\documentclass[11pt]{article}
\usepackage{graphicx}
\usepackage{amsmath}
\usepackage{amsfonts}
\usepackage{amssymb}

\newcommand{\be}{\begin{equation}}
\newcommand{\ee}{\end{equation}}
\newcommand{\tr}{{\rm Tr}}
\newcommand{\Wg}{{\rm Wg}}
\newcommand{\Z}{\mathcal{Z}}
\newcommand{\Y}{\mathcal{Y}}
\newcommand{\M}{\mathcal{M}}
\newcommand{\U}{\mathcal{U}}
\renewcommand{\O}{\mathcal{O}}

\begin{document}

\title{Expansion of polynomial Lie group integrals in terms of certain maps on surfaces, and factorizations of permutations}
\author{Marcel Novaes\\
{\small Instituto de F\'isica, Universidade Federal de Uberl\^andia, Uberl\^andia, MG,
38408-100, Brazil}}\date{} \maketitle
\begin{abstract}
Using the diagrammatic approach to integrals over Gaussian random matrices, we find a
representation for polynomial Lie group integrals as infinite sums over certain maps on
surfaces. The maps involved satisfy a specific condition: they have some marked vertices,
and no closed walks that avoid these vertices. We also formulate our results in terms of
permutations, arriving at new kinds of factorization problems.
\end{abstract}

\section{Introduction}

\subsection{Background}

We are interested in integrals over the orthogonal
group $\mathcal{O}(N)$ of $N$-dimensional real matrices $O$ satisfying $OO^T=1$, where
$T$ means transpose, and over the unitary group $\mathcal{U}(N)$ of $N$-dimensional complex
matrices $U$ satisfying $UU^\dag=1$, where $\dag$ means transpose conjugate. These groups have a unique invariant probability measure, known as Haar measure, and integrals over them may be seen
as averages over ensembles of random matrices. 

We will consider averages of functions that are polynomial in the matrix elements, i.e.
quantities like $ \prod_{k=1}^n
U_{i_k,j_k}U^\dag_{p_k,q_k}$ for the unitary group and
$ \prod_{k=1}^n O_{i_k,j_k}O_{i_{\widehat{k}},j_{\widehat{k}}}$ for the orthogonal one (results for the unitary symplectic group $Sp(N)$ are very close to those for $\mathcal{O}(N)$, so we do not consider this group in detail). From
the statistical point of view, these are joint moments of the matrix elements, considered
as correlated random variables. Their study started in physics \cite{weing,samuel}, with
applications to quantum chaos \cite{mello,BB96,esposti}, and found its way into
mathematics, initially for the unitary group \cite{collins} and soon after for the orthogonal
and symplectic ones \cite{ColSni,ColMat}. Since then, they have been extensively explored
\cite{zuber,Banica}, related to Jucys-Murphy elements \cite{MatNovak,zinn} and
generalized to symmetric spaces \cite{matsucompact}. Unsurprisingly, after these
developments some new applications have been found in physics
\cite{scott,pineda,cramer,znidaric,novaes1,novaes2}.

For the unitary group average, the result is different from zero only if the $q$-labels
are a permutation of the $i$-labels, and the $p$-labels are a permutation of the
$j$-labels. The basic building blocks of the calculation, usually called Weingarten
functions, are of the kind \be\label{weingU} \Wg^{U}_N(\pi)=\int_{\U(N)} dU
\prod_{k=1}^nU_{k,k}U^\dag_{k,\pi(k)},\ee where $\pi$ is an element of the permutation
group, $\pi\in S_n$. In general, if there is more than one permutation relating the sets
of labels (due to repeated indices, e.g. $\langle |U_{1,2}|^4\rangle$), the result is a
sum of Weingarten functions. The cycletype of a permutation in $S_n$ is a partition of
$n$, $\alpha=(\alpha_1,\alpha_2,...)\vdash n=|\alpha|$, whose parts are the lengths of
its cycles; the function $\Wg^{U}_N(\pi)$ depends only on the cycletype of $\pi$
\cite{collins,ColSni}.

The result of the orthogonal group average is different from zero only if the $i$-labels
satisfy some matching (see Section 2.1 for matchings) and the $j$-labels also satisfy some matching
\cite{ColSni,zuber,ColMat}. For concreteness, we may choose the $i$'s to satisfy only the trivial matching, and the $j$'s to satisfy only some matching $\mathfrak{m}$ of cosettype $\beta$. The basic building blocks of polynomial
integrals over the orthogonal group are \be\label{weingO} \Wg^{O}_{N}(\beta)=\int_{\O(N)}
dO O_{1,j_1}O_{1,j_{\widehat1}}O_{2,j_2}O_{2,j_{\widehat2}}\cdots O_{n,j_{n}}O_{n,j_{\widehat n}}.\ee 

As examples, let us mention \be\label{ex1} \Wg^{U}_N((2))=\int_{\U(N)} dU
U_{1,1}U^\dag_{1,2}U_{2,2}U^\dag_{2,1}=\frac{-1}{(N-1)N(N+1)},\ee corresponding to the
permutation $\pi=(12)$, which has cycletype $(2)$, and \be\label{ex2}\Wg^{O}_{N}((2))=\int_{\O(N)} dO
O_{1,1}O_{1,2}O_{2,2}O_{2,1}=\frac{-1}{(N-1)N(N+2)},\ee corresponding to the matching
$\{\{1,\widehat{2}\},\{2,\widehat 1\}\}$ for the $j$'s, which has cosettype $(2)$. 

Our subject is the combinatorics associated with the large $N$ asymptotics of these
integrals. For the unitary case, this has been addressed in previous works
\cite{BB96,collins,MatNovak,justin,guionnet}, where connections with maps and
factorizations of permutations have already appeared. However, our approach is different,
and the maps/factorizations we encounter are new. Our treatment of the
orthogonal case is also new. We proceed by first considering Weingarten functions in the context of
ensembles of random truncated matrices; then relating those to Gaussian ensembles, and
finally using the rich combinatorial structure of the latter.

In what follows we briefly present our results. A review of some basic concepts can be
found in Section 2. Results for the unitary group are obtained in Section 3, while the
orthogonal group is considered in Section 4. In an Appendix we discuss several different factorization problems that are 
related to $1/N$ expansions of Weingarten functions.

\subsection{Results and discussion for the unitary group}
\subsubsection{Maps}

For the unitary group, we represent the Weingarten function as an infinite sum over
orientable maps.

{\bf Theorem 1} \emph{If $\alpha$ is a partition with $\ell(\alpha)$ parts, then \be\label{result1}
\Wg^{U}_N(\alpha)=\frac{(-1)^{\ell(\alpha)}}{N^{2|\alpha|+\ell(\alpha)}}\sum_{\chi}N^{\chi}\sum_{w\in\mathcal{B}(\alpha,\chi)}
(-1)^{V(w)},\ee where the first sum is over Euler characteristic, and
$V(w)$ is the number of vertices in the map $w$.}

As we will discuss with more detail in Section 3.2 the (finite) set
$\mathcal{B}(\alpha,\chi)$ contains all maps, not necessarily connected, with the
following properties: i) they are orientable with Euler characteristic $\chi$; ii) they
have $\ell(\alpha)$ marked vertices with valencies ($2\alpha_1,2\alpha_2,...$); iii) all
other vertices have even valence larger than 2; iv) all closed walks along the boundaries
of the edges (which we see as ribbons) visit the marked vertices in exactly one corner;
v) they are face-bicolored and have $|\alpha|$ faces of each color.

\begin{figure}[tb]
\includegraphics[scale=0.8,clip]{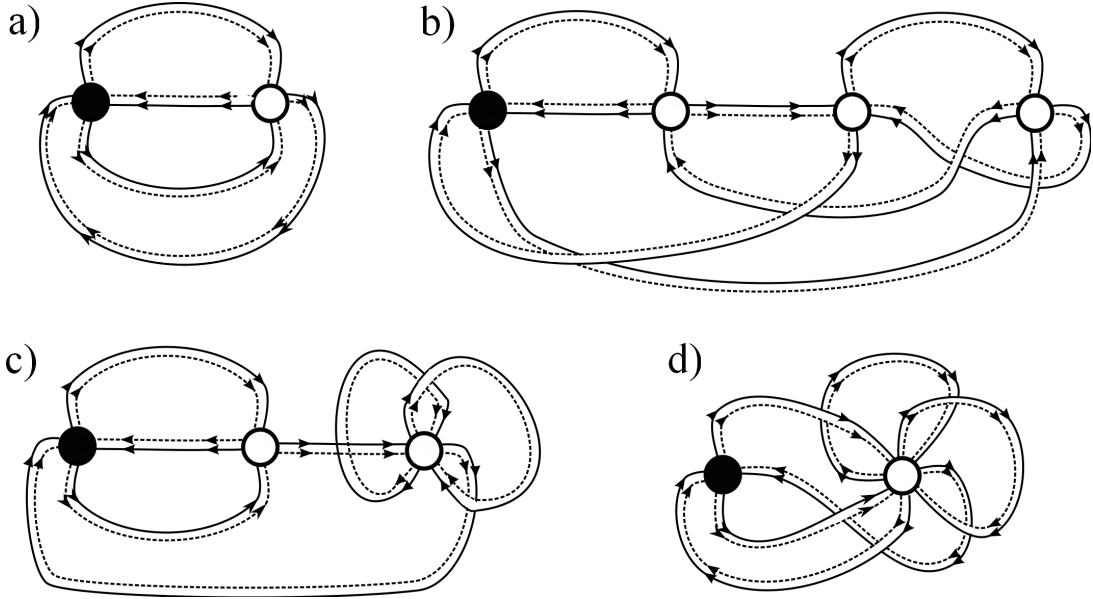}
\caption{A few of the maps that contribute to the Weingarten function (\ref{ex1}),
according to the expansion (\ref{result1}). They are face-bicolored, with different
colors being indicated by solid and dashed lines, respectively. All boundary walks visit
the marked vertex, which is the black disk. Their contributions are discussed in the
text.}
\end{figure}

As an example, the $1/N$ expansion for the function in (\ref{ex1}), for which
$\alpha=(2)$, starts \be\label{expU}-\frac{1}{N^3}-\frac{1}{N^5}-\frac{1}{N^7}-\cdots,\ee
The first two orders come from maps like the ones shown in Figure 1. The map in panel
$a)$ is the only leading order contribution. It has $\chi=2$ and $V=2$, so its value is
indeed $-1/N^3$. The next order comes from elements of $\mathcal{B}((2),0)$. There are in total 21 different maps in that set having four vertices of valence
$4$; one of them is shown in panel $b)$. There are 28 different maps in that set having two
vertices of valence $4$ and one vertex of valence $6$; one of them is shown in panel
$c)$. Finally, there are 8 different maps in that set having one vertex of valence $4$ and
one vertex of valence $8$; one of them is shown in panel $d)$. They all have $\chi=0$,
and their combined contribution is $(-21+28-8)/N^5=-1/N^5$.

\subsubsection{Factorizations}

By a factorization of $\Pi$ we mean an ordered pair, $f\equiv(\tau_1,\tau_2)$, such that
$\Pi=\tau_1\tau_2$. We call $\Pi$ the `target' of the factorization $f$. If $\Pi\in S_k$,
then the Euler characteristic of $f$ is given by
$\chi=\ell(\Pi)-k+\ell(\tau_1)+\ell(\tau_2)$.

The number of factorizations of a permutation depends only on its cycletype, so it makes
sense to restrict attention to specific representatives. Call a permutation `standard' if
each of its cycles contains only adjacent numbers, and the cycles have weakly decreasing
length when ordered with respect to least element. For example, $(123)(45)$ is standard.
Since $\Wg^{U}_N(\pi)$ depends only on the cycletype of $\pi$, we may take $\pi$ to be
standard.

As we will discuss with more detail in Section 3.3, the relevant factorizations for $\Wg^{U}_N(\pi)$ are those
whose target is of the kind $\Pi=\pi\rho$, where the `complement' $\rho$ is a standard
permutation acting on the set $\{n+1,...,n+m\}$, for some $m\ge 0$. They satisfy the following
properties: i) they have Euler characteristic $\chi$; ii) the complement $\rho$ has no
fixed points; iii) every cycle of the factors $\tau_1,\tau_2$ must have exactly one
element in $\{1,...,n\}$. Notice that the last condition implies
$\ell(\tau_1)=\ell(\tau_2)=n$. Let the (finite) set of all such factorizations be denoted
$\mathcal{F}(\pi,\chi)$. Then, we have

{\bf Theorem 2} \emph{Let $\pi\in S_n$ be a standard permutation, then
\be\label{result1b} \Wg^{U}_N(\pi)=\frac{(-1)^{\ell(\pi)}}{N^{2n+\ell(\pi)}}\sum_\chi
N^{\chi}\sum_{f\in\mathcal{F}(\pi,\chi)}\frac{(-1)^{\ell(\Pi)}}{z_\rho},\ee where $z_\rho=\prod_j j^{v_j}v_j!$,
with $v_j$ the number of times part $j$ occurs in the cycletype of the complement
$\rho$.}

Theorem 2 follows from Theorem 1 by a simple procedure for associating factorizations to
maps in $\mathcal{B}(\alpha,\chi)$, discussed in Section 3.3. Associations of this kind
are well known \cite{busquetmelou,irving,bouttier}.

For example, the leading order in Eq.(\ref{expU}), for which $\pi=(12)$, has a contribution from the
factorization $(12)(34)=(14)(23)\cdot (13)(24)$, which has $\rho=(34)$ and is 
one of two factorizations that can be associated with the map in Figure 1a. Several 
factorizations can be associated with the other maps in Figure 1. We mention one
possibility for each of them: $(12)(34)(56)(78)=(148)(25763)\cdot (13)(285746)$ for the
map in Figure 1b; $(12)(345)(67)=(17)(23546)\cdot (16)(27435)$ for the map in Figure 1c;
$(12)(3456)=(164)(253)\cdot(1365)(24)$ for the map in Figure 1d. Notice how they have
different complements.

Other factorization problems are also related to the coefficients in the $1/N$ expansion for $\Wg^U_N(\pi)$ (see the Appendix). 
Collins initially showed \cite{collins} that they can be expressed in terms of the number of `Proper' factorizations of $\pi$. Matsumoto and Novak
later showed \cite{MatNovak} that the coefficients count Monotone factorizations. On the other hand, Berkolaiko and Irving recently defined \cite{BerkoIrv} Inequivalent-Cycle factorizations and showed that
\be\label{equalities} \sum_{r\ge 0} (-1)^rI_{\alpha,\chi}(r)= \sum_{r\ge 0}
(-1)^rP_{\alpha,\chi}(r)=(-1)^{n+\ell(\alpha)}M_{\alpha,\chi}, \ee where
$I_{\alpha,\chi}(r)$ and $P_{\alpha,\chi}(r)$ are, respectively, the numbers of
Inequivalent-Cycle and Proper factorizations of $\pi$, with cycletype $\alpha\vdash n$,
into $r$ factors having Euler characteristic $\chi$, while $M_{\alpha,\chi}$ is the
number of Monotone factorizations of $\pi$ with Euler characteristic $\chi$.
Interestingly, our factorizations satisfy a very
similar sum rule, namely \be\label{sumrule}\sum_{f\in\mathcal{F}(\pi;\chi)}
\frac{(-1)^{\ell(\Pi)}}{z_\rho}=(-1)^{n+\ell(\alpha)}M_{\alpha,\chi}.\ee An important
difference between (\ref{sumrule}) and (\ref{equalities}) is that the factorizations in
(\ref{sumrule}) take place in $S_{n+m}$ for some $m\ge 0$, while all those in
(\ref{equalities}) take place in $S_n$.

Notice that our factorizations must satisfy condition iii), which is related to the
distribution of the elements from the set $\{1,...,n\}$ among the cycles of the factors
$\tau_1,\tau_2$. This is close in spirit to the kind of questions studied by B\'ona,
Stanley and others \cite{bona,stanley}, which count factorizations satisfying some
placement conditions on the elements of the target.

\subsection{Results and discussion for the orthogonal group}
\subsubsection{Maps}

For the orthogonal group, we represent the Weingarten function as an infinite sum over maps
(orientable and non-orientable).

{\bf Theorem 3} \emph{Let $\beta$ be a partition with $\ell(\beta)$ parts, then \be\label{result2}
\Wg^{O}_{N+1}(\beta)=\frac{(-2)^{\ell(\beta)}}{N^{2|\beta|+\ell(\beta)}}\sum_\chi
N^{\chi}\sum_{w\in\mathcal{NB}(\beta,\chi)}
\left(-\frac{1}{2}\right)^{V(w)},\ee where the first sum is over Euler characteristic,
and $V(w)$ is the number of vertices in the map $w$.}

\begin{figure}[tb]
\includegraphics[scale=0.8,clip]{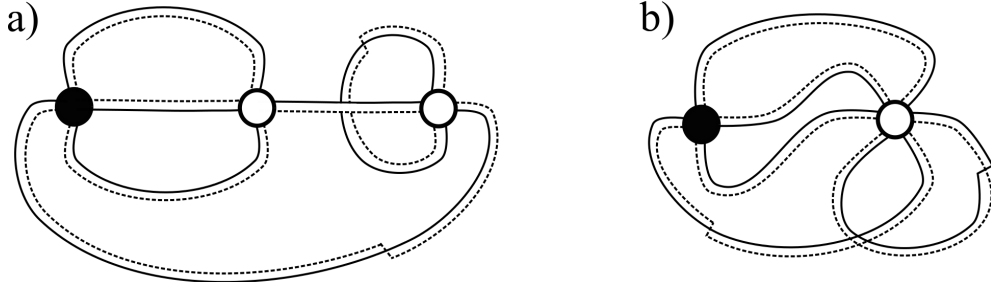}
\caption{Two of the maps that contribute to the Weingarten function (\ref{ex2}),
according to the expansion (\ref{result2}). They are not orientable: some of the ribbons
have `twists'.}
\end{figure}

As we will discuss with more detail in Section 4.2 the (finite) set
$\mathcal{NB}(\beta,\chi)$ contains all maps, not necessarily connected or orientable,
with the following properties: i) they have Euler characteristic $\chi$; ii) they have
$\ell(\beta)$ marked vertices with valencies ($2\beta_1,2\beta_2,...$); iii) all other
vertices have even valence larger than 2; iv) all closed walks along the boundaries of
the edges visit the marked vertices in exactly one corner; v) they are face-bicolored and
have $|\beta|$ faces of each color.

Notice that the expansion is in powers of $N^{-1}$ for the group $\mathcal{O}(N+1)$ and
not for $\mathcal{O}(N)$. For example, the first terms in the expansion of the function
in Eq.(\ref{ex2}) at dimension $N+1$ are
\be\label{expan2}\frac{-1}{N(N+1)(N+3)}=-\frac{1}{N^3}+\frac{4}{N^4}-\frac{13}{N^5}...\ee
The leading order comes from the map shown in Figure 1a, which is
orientable. The next order comes from non-orientable maps in the set $\mathcal{NB}((2),1)$. There are in
total 8 different maps having two vertices of valence $2$; one of them is shown in Figure
2a. There are in total 4 different maps having one vertex of valence $2$ and one vertex
of valence $3$; one of them is shown in Figure 2b. Their combined contribution is
$(8-4)/N^4=4/N^4$.

{\bf Remark} It is known \cite{matsucompact} that the Weingarten function of the unitary symplectic group $Sp(N)$ is proportional to $\Wg^{O}_{-2N}$. Therefore the appropriate dimension for map expansion in $Sp(N)$ is also different from $N$: it
has to be $Sp(N-1/2)$ (a non-integer dimension is to be understood by keeping in mind
that Weingarten functions are rational functions of $N$). Interestingly, if we assign a parameter
$\alpha=2,1,1/2$ for orthogonal, unitary and symplectic groups, respectively (sometimes
called Jack parameter), then the appropriate dimensions for the map expansion in powers
of $N^{-1}$ can be written as $N+\alpha-1$.

\subsubsection{Factorizations}

Let $[n]=\{1,...,n\}$ and $\widehat{[n]}=\{\widehat1,...,\widehat n\}$. We consider the action of $S_{2n}$ on $[n]\cup\widehat{[n]}$. Define the
`hat' involution on permutations as $\widehat\pi(a)=\pi^{-1}(\widehat a)$ (assuming $\widehat{\widehat a}=a$). We
call permutations that are invariant under this transformation `palindromic', e.g.
$(12\widehat 2 \widehat 1)$ and $(12)(\widehat 2 \widehat 1)$ are palindromic. 

Given a partition $\beta$, define $\pi\in S_n$ to be the standard permutation that has cycletype $\beta$ and define
$\Pi=\pi\rho$, where the `complement' $\rho$ is a standard permutation acting on the set $\{n+1,...,n+m\}$ for some $m\ge 0$. Define the fixed-point free involutions $p_1$, whose cycles are $(a\,\widehat{a})$ for $1\le a\le n+m$, and $p_2$ whose cycles are of the type $(\widehat{a}\,a+1)$, but with the additions computed modulo the cycle lengths of $\Pi$, i.e. \be
p_2=(\widehat1\,2)(\widehat2\,3)\cdots(\widehat{\beta_1}\,1)(\widehat{\beta_1+1}\,\beta_1+2)\cdots(\widehat{\beta_1+\beta_2}\,\beta_1+1)\cdots.\ee
They provide a factorization of the palindromic version of the target, $p_2p_1=\Pi\,\widehat{\Pi}$. 

The problem we need to solve is to find all factorizations $\Pi\,\widehat{\Pi}=f_2 f_1$, that satisfy the following properties: i) their Euler characteristic, defined as $\ell(\Pi)-m-n+\ell(f_1)+\ell(f_2)$, is $\chi$; ii) the complement $\rho$ has no fixed points; iii) the factors may be written as $f_1=\theta p_1$ and $f_2=p_2\theta$ for some fixed-point free involution $\theta$; iv) $f_1$ is palindromic; v) every cycle of the factors $f_1,f_2$ contains exactly one element in $[n]\cup\widehat{[n]}$. Clearly, the crucial quantity is actually $\theta$. Let the
(finite) set of all pairs $(\Pi,\theta)$ satisfying these conditions be denoted
$\mathcal{NF}(\beta,\chi)$. Then, we have

{\bf Theorem 4} \emph{For a given partition $\beta$ of length $\ell(\beta)$, 
\be
\Wg^{O}_{N+1}(\beta)=\frac{(-2)^{\ell(\beta)}}{N^{2|\beta|+\ell(\beta)}}\sum_\chi
N^{\chi}\sum_{(\Pi,\theta)\in\mathcal{NF}(\beta,\chi)}\frac{1}{z_\rho}\left(-\frac{1}{2}\right)^{\ell(\Pi)},\ee
where $z_\rho=\prod_j j^{v_j}v_j!$, with $v_j$ the number of times part $j$ occurs in the
cycletype of the complement $\rho$.}

Theorem 4 follows from Theorem 3 by a simple procedure for describing combinatorially the maps in
$\mathcal{NB}(\beta,\chi)$, discussed in Section 4.3. Such kind of descriptions are well
known \cite{vassili,hanlon,jackson}.

For example, the leading order in Eq.(\ref{expan2}) has a contribution from the pair
$\Pi=(12)(34)$, $\theta=(1\widehat3)(2\widehat4)(4\widehat2)(3\widehat1)$. The next order comes from factorizations associated with the maps in Figure 2. We mention one for each of them: $\Pi=(12)(34)(56)$, $\theta=(1\widehat3)(\widehat13)(25)(\widehat24)(\widehat46)(\widehat5\widehat6)$ for a) and
$\Pi=(12)(345)$, $\theta=(1\widehat3)(\widehat13)(25)(\widehat24)(\widehat4\widehat5)$ for b).

Other factorizations are related to the coefficients in the $1/N$ expansion of orthogonal Weingarten functions, as we discuss in the Appendix. Matsumoto has shown \cite{matsujack} that these coefficients count certain factorizations that are matching analogues of the Monotone family. Following the steps in \cite{collins} we show that the coefficients can be expressed in terms of analogues of Proper factorizations. These relations hold for $\mathcal{O}(N)$, however, not $\mathcal{O}(N+1)$. Therefore the relation to our results is less direct. On the other hand, Berkolaiko and Kuipers have shown \cite{GregJack1} that certain `Palindromic Monotone' factorizations are related to the $1/N$ expansion for $\mathcal{O}(N+1)$. The appropriate analogue of Inequivalent-Cycle factorizations is currently missing. 

\subsection{Connection with physics}

This work was originally motivated by applications in physics, in the semiclassical
approximation to the quantum mechanics of chaotic systems. Without going into too much
detail, the problem in question was to construct correlated sets of chaotic trajectories that are
responsible for the relevant quantum effects in the semiclassical limit \cite{transp}. Connections
between this topic and factorizations of permutations had already been noted
\cite{haake1,GregJack1,GregJack2,epl,combprob}. In \cite{novaes1} we suggested that
such sets of trajectories could be obtained from the diagrammatic expansion of a certain
matrix integral, with the proviso that the dimension of the matrices had to be set to
zero after the calculation. When we realized the connection to truncated random unitary
matrices, this became our Theorem 1. The suggestion from \cite{novaes1}, initially
restricted to systems without time-reversal symmetry, was later extended to systems that
have this symmetry \cite{novaes2}; the connection with truncated random orthogonal
matrices then gave rise to our Theorem 3.

\subsection{Acknowledgments}

The connection between our previous work \cite{novaes1} and truncated unitary matrices was first
suggested by Yan Fyodorov.

This work had financial support from CNPq (PQ-303318/2012-0) and FAPEMIG (APQ-00393-14).

\section{Basic concepts}

\subsection{Partitions, Permutations and Matchings}

By $\alpha\vdash n$ we mean $\alpha$ is a partition of $n$, i.e. a weakly decreasing
sequence of positive integers such that $\sum_i \alpha_i=|\alpha|=n$. The number of
non-zero parts is called the length of the partition and denoted by $\ell(\alpha)$.

The group of permutations of $n$ elements is $S_n$. The cycletype of $\pi\in S_n$ is the
partition $\alpha\vdash n$ whose parts are the lengths of the cycles of $\pi$. The length
of a permutation is the number of cycles it has, $\ell(\pi)=\ell(\alpha)$, while its rank is $r(\pi)=n-\ell(\pi)=r(\alpha)$. We multiply
permutations from right to left, e.g. $(13)(12)=(123)$. The conjugacy class $C_\lambda$ is the set of all permutations with cycletype $\lambda$. Its size is $|C_\lambda|=n!/z_\lambda$, with $z_\lambda=\prod_j j^{v_j}v_j!$, where $v_j$ is the number of times part $j$ occurs in $\lambda$.

Let $[n]=\{1,...,n\}$, $\widehat{[n]}=\{\widehat1,...,\widehat n\}$ and let $[n]\cup\widehat{[n]}$ be a set with $2n$ elements. A \emph{matching} on this set is a collection of $n$ disjoint subsets with two elements each. The set of all such matchings is $\mathcal{M}_n$. The \emph{trivial} matching is defined as $\mathfrak{t}=\{\{1,\widehat1\},\{2,\widehat2\},...,\{n,\widehat n\}\}$. A permutation $\pi$ acts on matchings by replacing blocks like $\{a,b\}$ by $\{\pi(a),\pi(b)\}$. If $\pi(\mathfrak{t})=\mathfrak{m}$ we say that $\pi$ \emph{produces} the matching $\mathfrak{m}$.

Given a matching $\mathfrak{m}$, let
$\mathcal{G}_\mathfrak{m}$ be a graph with $2n$ vertices having labels in $[n]\cup\widehat{[n]}$, two vertices
being connected by an edge if they belong to the same block in either $\mathfrak{m}$ or $\mathfrak{t}$. Since each vertex belongs to two edges, all connected components of
$\mathcal{G}_\mathfrak{m}$ are cycles of even length. The \emph{cosettype} of $\mathfrak{m}$ is the
partition of $n$ whose parts are half the number of edges in the connected components of
$\mathcal{G}_\mathfrak{m}$. See examples in Figure 3. A permutation has the same cosettype as the matching it produces.

\begin{figure}[tb]
\includegraphics[scale=1,clip]{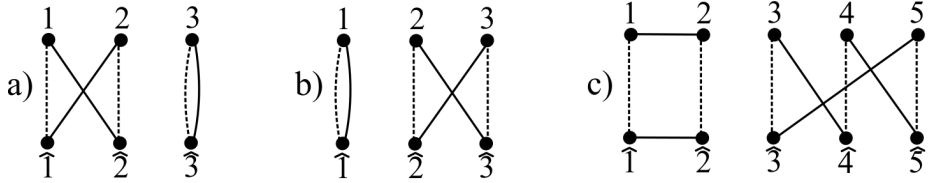}
\caption{Three examples of graphs associated with matchings. Edges coming from the trivial matching, $\mathfrak{t}$, are drawn in dashed line. In a) we have $\mathfrak{m}=\{\{1,\widehat2\},\{2,\widehat1\},\{3,\widehat3\}\}$, which can be produced as $(12)(\mathfrak{t})$ or $(\widehat1\widehat2)(\mathfrak{t})$, among others ways; permutations $(12)$ and $(\widehat1\widehat2)$ thus belong to the same coset in $S_{2n}/H_n$. In b) $\mathfrak{m}=\{\{2,\widehat3\},\{3,\widehat2\}\}$, which can be produced as $(23)(\mathfrak{t})$ or $(\widehat2\widehat3)(\mathfrak{t})$, among others. The matchings in both a) and b) have the same cosettype, namely $(2,1)$; the permutations producing them therefore belong to the same double coset in $H_n\backslash S_{2n}/H_n$. In c) 
$\mathfrak{m}=\{\{1,2\},\{\widehat1,\widehat2\},\{3,\widehat4\},\{4,\widehat5\},\{5,\widehat3\}\}$, and its cosettype is
$(3,2)$.}
\end{figure}

We realize the group $S_{2n}$ as the group of all permutations acting on the set $[n]\cup\widehat{[n]}$. It has a subgroup called the hyperoctahedral, $H_n$, with $|H_n|=2^nn!$
elements, which leaves invariant the trivial matching and is generated by permutations of the form $(a\,\widehat{a})$ or of the form $(a\,b)(\widehat{a}\,\widehat{b})$. The cosets $S_{2n}/H_n\sim\M_n$ can therefore be represented by matchings. The trivial matching identifies the coset of the identity permutation. We may inject $\mathcal{M}_n$ into $S_{2n}$ by using fixed-point free involutions. This is done by the simple identification $\mathfrak{m}=\{\{a_1,a_2\},\{a_3,a_4\},...\} \mapsto \sigma_\mathfrak{m}=
(a_1\,a_2)(a_3\,a_4)\cdots$. 

The double cosets $H_n\backslash S_{2n}/H_n$, on the other hand, are indexed by partitions of $n$: two permutations belong to the same double coset if and only if they have the same cosettype \cite{donald} (hence this terminology). We denote by $K_\lambda$ the double coset of all permutations with cosettype $\lambda$. Its size is $|K_\lambda|=|H_n||C_\lambda|2^{r(\lambda)}$. 

Given a sequence of $2n$ numbers, $(i_1,...,i_{2n})$ we say that it satisfies the
matching $\mathfrak{m}$ if the elements coincide when
paired according to $\mathfrak{m}$. This is quantified by the function $\Delta_\mathfrak{m}(i)=\prod_{b\in\mathfrak{m}}\delta_{i_{b_1},i_{b_2}}$, where the product runs
over the blocks of $\mathfrak{m}$ and $b_1,b_2$ are the elements of block $b$. 

\subsection{Maps}

Maps are graphs drawn on a surface, with a well defined sense of rotation around each
vertex. We represent the edges of a map by ribbons. These ribbons meet at the vertices,
which are represented by disks, and the region where two ribbons meet is called a
\emph{corner}.

It is possible to go from one vertex to another by walking along a boundary of a ribbon.
Upon arriving at a vertex, a walker may move around the boundary of the disk to a
boundary of the next ribbon, and then depart again. Such a walk we call a \emph{boundary walk}. A boundary walk
that eventually retraces itself is called \emph{closed} and delimits a
\emph{face} of the map.

Let $V$, $E$ and $F$ be respectively the numbers of vertices, edges and faces of a map.
The Euler characteristic of the map is $\chi=V-E+F$, and it is additive in the connected
components (we do not require maps to be connected, and each connected component is
understood to be drawn on a different surface).

In all of the maps used in this work, ribbons have boundaries of two different colors. For convenience, we represent those colors by simply drawing these boundaries with two types os lines: dashed lines and solid lines. Ribbons are attached to vertices in such a way that all corners and all faces have a well defined color, i.e. our maps are \emph{face-bicolored}. Examples are shown in Figures 1 and 2. 

\subsection{Non-hermitian Gaussian random matrices}

We consider $N$-dimensional matrices for which the elements are independent and identically-distributed
Gaussian random variables, the so-called Ginibre ensembles \cite{ginibre}.
For real matrices we will use the notation $M$, for complex ones we use $Z$.

\begin{figure}[tb]\label{Wick}
\includegraphics[scale=0.8,clip]{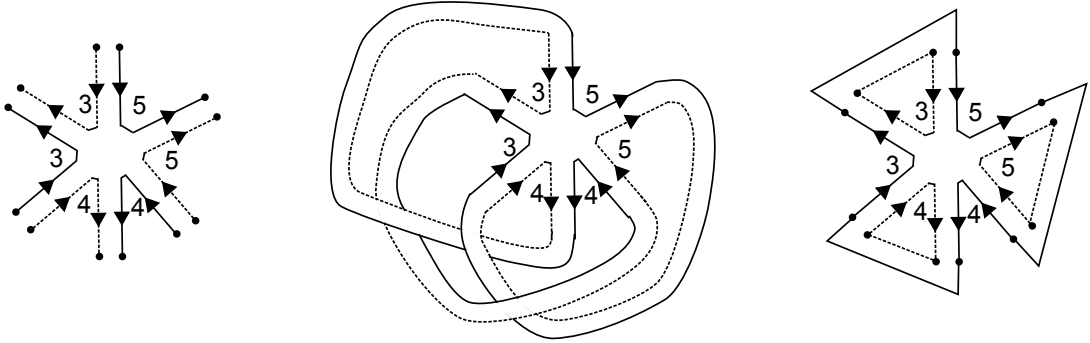}
\caption{Wick's rule for complex Gaussian random matrices. Starting from $\langle {\rm
Tr}(ZZ^\dag)^3\rangle=\sum_{a_1,a_2,a_3}\sum_{b_1,b_2,b_3}Z_{a_1b_1}Z^\dag_{b_1a_2}Z_{a_2b_2}Z^\dag_{b_2a_3}Z_{a_3b_3}Z^\dag_{b_3a_1}$,
we arrange the elements around a vertex (left panel). Then, we produce all possible
connections between the marked ends of the arrows, respecting orientation. Two of them are shown. The map in the
middle panel has only one face of each color, and $\chi=0$. The map in the right panel
has three faces of one color and a single face of the other, with $\chi=2$.}
\end{figure}

Normalization constants for these ensembles are defined as \be \Z_R=\int dM
e^{-\frac{\Omega}{2} {\rm Tr}(MM^T)}, \quad
 \Z_C=\int dZ e^{-\Omega {\rm Tr}(ZZ^\dag)}.\ee
They can be computed (as we do later) using singular value decomposition. Average values
are denoted by \be\langle f(M)\rangle= \frac{1}{\Z_R}\int dMe^{-\frac{\Omega}{2} {\rm
Tr}(MM^T)}f(M),\ee and \be\langle f(Z)\rangle= \frac{1}{\Z_C}\int dZe^{-\Omega {\rm
Tr}(ZZ^\dag)}f(Z),\ee the meaning of $\langle \cdot \rangle$ being clear from context. We
have the simple covariances \be\label{covr} \langle M_{ab}M_{cd}\rangle=\frac{1}{\Omega}\delta_{ac}\delta_{bd},\ee and \be\label{covc} \langle Z_{ab}Z_{cd}\rangle=\langle Z_{ab}^*Z^*_{cd}\rangle=0, \quad
\langle Z_{ab}Z^*_{cd}\rangle=\frac{1}{\Omega}\delta_{ac}\delta_{bd}.\ee 

\begin{figure}[tb]\label{WickO}
\includegraphics[scale=0.8,clip]{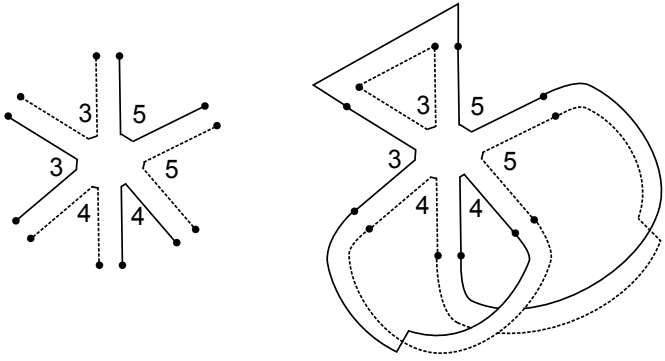}
\caption{Wick's rule for real Gaussian random matrices. Starting from $\langle {\rm
Tr}(MM^T)^3\rangle$, we arrange the elements around a vertex (left panel). Then, we
produce all possible connections between the marked ends of the arrows. One of them is
shown, which has two faces of one color and a single face of the other, with $\chi=1$.
Notice that the boundaries of the edges are not oriented, and that two of them have
twists.}
\end{figure}

Polynomial integrals may be computed using Wick's rule, which is a combinatorial
prescription for combining covariances. It simply states that, since the elements are
independent, the average of a product can be decomposed in terms of products of
covariances. In the complex case, we may consider the elements of $Z$ fixed and then
permute the elements of $Z^\dag$ in all possible ways,  \be \left\langle\prod_{k=1}^n
Z_{a_kb_k}Z^\dag_{c_{k}d_k}\right\rangle=\frac{1}{\Omega^n}\sum_{\pi\in S_n}\prod_{k=1}^n
\delta_{a_k,d_{\pi(k)}}\delta_{b_k,c_{\pi(k)}}.\ee For example, \be\label{exZ} \langle
Z_{a_1b_1}Z^*_{d_1c_1}Z_{a_2b_2}Z^*_{d_2c_2}\rangle=
\frac{1}{\Omega^2}\delta_{a_1,d_{1}}\delta_{b_1,c_{1}}\delta_{a_2,d_{2}}\delta_{b_2,c_{2}}
+\frac{1}{\Omega^2}\delta_{a_1,d_{2}}\delta_{b_1,c_{2}}\delta_{a_2,d_{1}}\delta_{b_2,c_{1}}.
\ee In the real case, we must consider all possible matchings among the
elements, \be \left\langle\prod_{k=1}^{2n}
M_{a_kb_k}\right\rangle=\frac{1}{\Omega^n}\sum_{\mathfrak{m}\in
\mathcal{M}_n}\Delta_\mathfrak{m}(a)\Delta_\mathfrak{m}(b).\ee For
example,\be\label{exM}\begin{alignedat}{2} \langle
M_{a_1b_1}M_{a_2b_2}M_{a_3b_3}M_{a_4b_4}\rangle&=
\frac{1}{\Omega^2}\delta_{a_1,a_2}\delta_{a_3,a_4} \delta_{b_1,b_2}\delta_{b_3,b_4}
+\frac{1}{\Omega^2}\delta_{a_1,a_3}\delta_{a_2,a_4}
\delta_{b_1,b_3}\delta_{b_2,b_4}\\&+\frac{1}{\Omega^2}\delta_{a_1,a_4}\delta_{a_2,a_3}
\delta_{b_1,b_4}\delta_{b_2,b_3}.
\end{alignedat}\ee

These Wick's rules have a well known diagrammatic interpretation (see, e.g.,
\cite{morris,hooft,bessis,fran,time}). In the complex case, matrix elements
are represented by ribbons having borders oriented in the same direction but with
different colors. Ribbons from elements of $Z$ have a marked head, while ribbons from
elements of $Z^\dag$ have a marked tail. Ribbons coming from traces are arranged around
vertices, so that all marked ends are on the outside and all corners have a well defined
color. Wick's rule consists in making all possible connections between ribbons, using
marked ends, respecting orientation. This produces a map (not necessarily connected). According to
Eq.(\ref{covc}), each edge leads to a factor $\Omega^{-1}$. In the real
cases, the boundaries of the ribbons are not oriented and the maps need not be
orientable: the edges may contain a `twist'. We show an example for the complex case in
Figure 4, and an example for the real case in Figure 5.

\section{Unitary Group}

\subsection{Truncations}

Let $U$ be a random matrix uniformly distributed in $\mathcal{U}(N)$ with the appropriate
normalized Haar measure. Let $A$ be the $M_1\times M_2$ upper left corner of $U$, with
$N\ge M_1+M_2$ and $M_1\le M_2$. It is known
\cite{truncU1,truncU2,truncU3,truncU5,truncU6} that $A$, which satisfies $AA^\dag<
1_{M_1}$, becomes distributed with probability density given by \be
P(A)=\frac{1}{\Y_1}\det(1-AA^\dag)^{N_0},\ee where  \be N_0=N-M_1-M_2\ee and $ \Y_1$ is a
normalization constant.

The value of $\Y_1$ can be computed using the singular value decomposition $A=WDV$, where
$W$ and $V$ are matrices from $\mathcal{U}(M_1)$ and $\mathcal{U}(M_2)$, respectively.
Matrix $D$ is real, diagonal and non-negative. Let $T=D^2={\rm diag}(t_1,t_2,...)$. Then
\cite{morris,shen,edelman}, \be
\Y_1=\int_{\mathcal{U}(M_1)}dW\int_{\mathcal{U}(M_2)}dV\int_0^1
\prod_{i=1}^{M_1}dt_i(1-t_i)^{N_0}t_i^{M_2-M_1}|\Delta(t)|^{2} ,\ee where \be
\Delta(t)=\prod_{i=1}^{M_1}\prod_{j=i+1}^{M_1}(t_j-t_i).\ee If we denote the angular
integrals by \be \int_{\mathcal{U}(M_1)}dW\int_{\mathcal{U}(M_2)}dV=\mathcal{V}_1,\ee
then Selberg's integral tells us that \cite{Selb1,Selb2}\be
\Y_1=\mathcal{V}_1\prod_{j=1}^{M_1}\frac{\Gamma(j+1)\Gamma(M_2+1-j)
\Gamma(N-M_2-j+1)}{\Gamma(N-j+1)}.\ee

Consider now an even smaller subblock of $U$, which is contained in $A$. Namely, let
$\widetilde{U}$ be the $N_1\times N_2$ upper left corner of $U$, with $N_1\le M_1$ and
$N_2\le M_2$. We shall make use of the obvious fact that integrals involving matrix
elements of $\widetilde{U}$ can be computed either by integrating over $U$ or over $A$.
In particular, the quantity \be\label{I1U} \Wg^{U}_N(\pi)=\int_{\mathcal{U}(N)} dU
\prod_{k=1}^n \widetilde{U}_{k,k}\widetilde{U}^\dag_{k,\pi(k)},\ee with $n\leq N_1,N_2$
and $\pi\in S_n$, can also be written as
\be\label{I1}\Wg^{U}_N(\pi)=\frac{1}{\Y_1}\int_{AA^\dag< 1_{M_1}} dA\det(1-AA^\dag)^{N_0}
\prod_{k=1}^n A_{k,k}A^\dag_{k,\pi(k)}.\ee

Notice that, although this may not be evident at first sight, the right-hand-side of
equation (\ref{I1}) is actually independent of $M_1$ and $M_2$.

\subsection{Sum over maps}

The key to the diagrammatic formulation of our integral is the identity \be\label{det2tr}
\det(1-AA^\dag)^{N_0}=e^{N_0{\rm Tr
log}(1-AA^\dag)}=e^{-N_0\sum_{q=1}^\infty\frac{1}{q}{\rm Tr}(AA^\dag)^q}.\ee We shall
consider the first term in the series separately from the rest, and incorporate it into
the measure, i.e. we will write \be dAe^{-N_0\sum_{q=1}^\infty\frac{1}{q}{\rm
Tr}(AA^\dag)^q}=d_G(A)e^{-N_0\sum_{q=2}^\infty\frac{1}{q}{\rm Tr}(AA^\dag)^q},\ee where
$d_G(A)$ is a Gaussian measure, \be d_G(A)=dAe^{-N_0{\rm Tr}(AA^\dag)}.\ee

We have \be \Wg^{U}_N(\pi)=\frac{1}{\Y_1}\int_{AA^\dag< 1_{M_1}}
d_G(A)e^{-N_0\sum_{q=2}^\infty\frac{1}{q}\tr(AA^\dag)^q}\prod_{k=1}^n
A_{k,k}A^\dag_{k,\pi(k)}.\ee Taking into account that the series in the exponent diverges
for $AA^\dag\geq 1_{M_1}$ and that $e^{-\infty}=0$, we extend the integration to general
matrices $A$, \be\label{Iabcd} \Wg^{U}_N(\pi)=\frac{1}{\Y_1}\int
d_G(A)e^{-N_0\sum_{q=2}^\infty\frac{1}{q}{\rm Tr}(AA^\dag)^q} \prod_{k=1}^n
A_{k,k}A^\dag_{k,\pi(k)}.\ee

Now we are in the realm of Gaussian integrals, and may apply Wick's rule. For each cycle
of $\pi$, the elements of $A$ and $A^\dag$ in the last product can be arranged in
counterclockwise order around vertices. This produces what we call `marked' vertices, in number
of $\ell(\pi)$. Formally expanding the exponential in Eq.(\ref{Iabcd}) as a Taylor series in
$N_0$ will produce other vertices, let us call them `internal', all of them of even
valence larger than 2. This leads to an infinite sum over maps with arbitrary numbers of internal vertices and
edges. The contribution of a map will be proportional to $(-1)^vN_0^{v-E}$,
if it has $v$ internal vertices and $E$ edges, which can be written as \be(-1)^vN_0^{v-E}=\frac{(-1)^{\ell(\pi)}}{N_0^{F+\ell(\pi)}}N_0^\chi(-1)^V.\ee 

However, the application of Wick's rule may lead to closed boundary walks that visit only
the internal vertices and avoid the marked ones. If a map has $r_1$ closed boundary walks
of one color and $r_2$ closed boundary walks of the other color that avoid the marked
vertices, its contribution will be proportional to $M_1^{r_1}M_2^{r_2}$. Crucially, since
we know that $\Wg^{U}_N(\pi)$ is actually independent of $M_1,M_2$, we are free to consider only the cases $r_1=r_2=0$, or equivalently to take
the simplifying limits $M_1\to 0$ and $M_2\to 0$. We will therefore be left only with
maps in which all closed boundary walks visit the marked vertices.

Another point we must address is that the normalization constant $\Y_1$ is not equal to
the corresponding Gaussian one, $\Z_C=\int d_G(A)$, so that the diagrammatic expression
for $ \Wg^{U}_N(\pi)$ will be multiplied by $\Z_C/\Y_1$. Using again the singular value
decomposition of $A$, and \be\label{SelbergE} \int_0^\infty
\prod_{i=1}^{M_1}dt_ie^{-N_0t_i}t_i^{M_2-M_1}|\Delta(t)|^{2}
=\left(\frac{1}{N_0}\right)^{M_1M_2}\prod_{j=1}^{M_1}\Gamma(j+1)\Gamma(M_2+1-j),\ee which
can be obtained as a limiting case of the Selberg integral, we arrive at \be
\Z_C=
\mathcal{V}_1\frac{1}{N_0^{M_1M_2}}\prod_{j=1}^{M_1}j!(M_2-j)!\ee and, therefore, \be
\frac{\Z_C}{\Y_1}=\frac{1}{N_0^{M_1M_2}}\prod_{j=1}^{M_1}\frac{(N-j)!}{(N-M_2-j)!}.\ee In
the end, we have the simplification\be \lim_{M_2\to 0}\frac{\Z_C}{\Y_1}=1.\ee The limit
$M_1\to 0$ is trivial. Notice that the limit of $N_0$ is simply $N$.

Because of the way we arranged the elements around the marked edges, our maps will always
have $2n$ faces, being $n$ of each color. We have therefore produced the maps in $\mathcal{B}(\alpha,\chi)$ and proved our Theorem 1.

\subsection{Sum over factorizations}

Suppose a standard permutation $\pi\in S_n$ with cycletype $\alpha\vdash n$. We shall
associate factorizations of $\pi$ with the maps in $\mathcal{B}(\alpha,\chi)$. In order to do that, we must label the corners of the maps. We proceed as follows.

Consider first the marked vertex with largest valency, and pick a corner delimited by solid line. Label this corner and the corner following it in counterclockwise order with number $1$. Then label the next two corners with the number $2$, and so on until
the label $\alpha_1$ is used twice. Proceed to another marked vertex in weakly
decreasing order of valency, and repeat the above with integers from
$\alpha_1+1$ to $\alpha_1+\alpha_2$. Repeat until all marked vertices have been labelled,
producing thus the cycles of $\pi$. The same procedure is then applied to the internal
vertices, producing the cycles of another standard permutation $\rho$, acting on the set
$\{n+1,...,E\}$, where $E=n+m$ is the number of edges in the map. Notice that $\rho$ has no fixed points, since all internal vertices of maps in $\mathcal{B}(\alpha,\chi)$ have even valence larger than 2.

Let $\Pi=\pi\rho$. See Figure 6, where $\Pi=(12)(34)$
for panel $a)$, $\Pi=(12)(34)(56)(78)$ for panel $b)$, $\Pi=(12)(345)(67)$ for panel
$c)$, and $\Pi=(12)(3456)$ for panel $d)$.

\begin{figure}[tb]\label{figX}
\includegraphics[scale=0.8,clip]{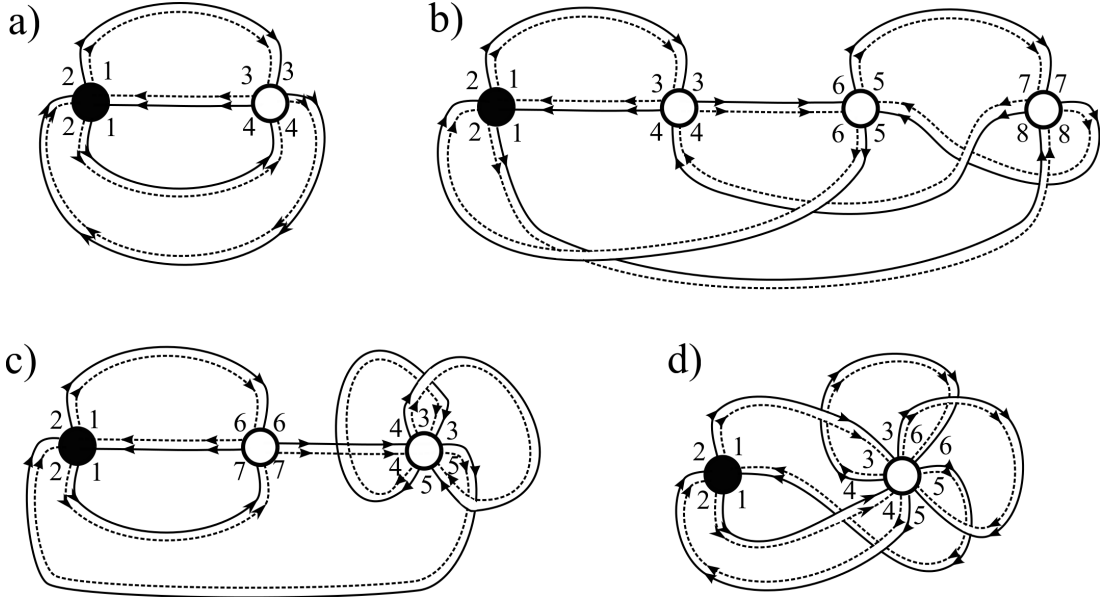}
\caption{Labeling the maps from Figure 1, we can associate with them certain
factorizations of permutations.}
\end{figure}

Define the permutation $\omega_1$ to be such that its cycles have the integers in the
order they are visited by the arrows in solid line. In the example of Figure 6,
this would be $\omega_1=(14)(23)$ for panel $a)$, $\omega_1=(184)(23675)$ for panel $b)$,
$\omega_1=(17)(26453)$ for panel $c)$ and $\omega_1=(146)(235)$ for panel $d)$. The cycles of $\omega_1$ correspond to the closed boundary walks in solid line.

Permutation $\tau_2$ is defined analogously in terms of the arrows in
dashed lines. In Figure 6, this would be $\tau_2=(13)(24)$ for panel $a)$,
$\tau_2=(13)(285746)$ for panel $b)$, $\tau_2=(16)(27435)$ for panel $c)$ and
$\tau_2=(1365)(24)$ for panel $d)$. The cycles of $\tau_2$ correspond to
the closed boundary walks in dashed line.

Suppose an initial integer, $i$. The arrow in dashed line which departs from the corner
labelled by $i$ arrives at the corner labelled by $\tau_2(i)$. On the other hand, the
image of $i$ under the permutation $\Pi$ corresponds to the label of an outgoing arrow in
solid line which, following $\omega_1$, also arrives at $\tau_2(i)$. Therefore, we have, by construction, $\omega_1\Pi=\tau_2$ or, equivalently, writing
$\tau_1=\omega_1^{-1}$, we have the factorization \be\label{factoriza}
\Pi=\tau_1\tau_2.\ee

For the maps in
$\mathcal{B}(\alpha,\chi)$ all boundary walks visit the marked vertices, which means that all
cycles of $\tau_1$ ad of $\tau_2$ must have exactly one element in the set $\{1,...,n\}$.
Therefore, the permutations satisfy the conditions we listed in Section 1.2.2. 

When we label the vertices of the map to produce a factorization, there are two kinds of
ambiguities. First, for a vertex of valency $2j$ there are $j$ possible choices for the
first corner to be labelled. Second, if there are $m_j$ vertices of valency $j$, there
are $m_j!$ ways to order them. Hence, to a map for which the complement is $\rho$ there correspond $z_\rho=\prod_j j^{m_j}m_j!$ factorizations, where $m_j$ is
the multiplicity of part $j$ in the cycletype of $\rho$. The sum in (\ref{result1}) can indeed be
written as \be \Wg^{U}_N(\pi)=\frac{1}{N^{2n+\ell(\pi)}}\sum_\chi
N^{\chi}\sum_{f\in\mathcal{F}(\pi,\chi)} \frac{(-1)^{\ell(\rho)}}{z_\rho},\ee
where $\mathcal{F}(\pi,\chi)$ is the set of factorizations of the kind we have
described for given $\alpha$ and $\chi$. This
proves our Theorem 2.

\section{Orthogonal group}

\subsection{Truncations}

Let $O$ be a random matrix uniformly distributed in $\mathcal{O}(N+1)$ with the
appropriate normalized Haar measure. Let $A$ be the $M_1\times M_2$ upper left corner of
$O$, with $N\ge M_1+M_2$ and $M_1\le M_2$. It is known \cite{truncU5,truncO} that $A$,
which satisfies $AA^T< 1_{M_1}$, becomes distributed with probability density given by
\be P(A)=\frac{1}{\Y_2}\det(1-AA^T)^{N_0/2},\ee where  \be N_0=N-M_1-M_2\ee and $ \Y_2$
is a normalization constant. Notice that we start with $\mathcal{O}(N+1)$ and not
$\mathcal{O}(N)$.

The value of $\Y_2$ can be computed using the singular value decomposition $A=WDV$, where
$W$ and $V$ are matrices from $\mathcal{O}(M_1)$ and $\mathcal{O}(M_2)$, respectively.
Matrix $D$ is real, diagonal and non-negative. Let $T=D^2={\rm diag}(t_1,t_2,...)$. Then
\cite{shen,edelman}, \be \Y_2=\int_{\mathcal{O}(M_1)}dW\int_{\mathcal{O}(M_2)}dV\int_0^1
\prod_{i=1}^{M_1}dt_i(1-t_i)^{N_0/2}t_i^{(M_2-M_1-1)/2}|\Delta(t)| .\ee If we denote the
angular integrals by $\int_{\mathcal{O}(M_1)}dW\int_{\mathcal{O}(M_2)}dV=\mathcal{V}_2,$
then we have again from Selberg's integral that \be
\Y_2=\mathcal{V}_2\prod_{j=1}^{M_1}\frac{\Gamma(j/2+1)\Gamma((M_2+1-j)/2)
\Gamma((N-M_2-j)/2+1)}{\Gamma((N-j)/2+1)\Gamma(3/2)}.\ee

Consider now an even smaller subblock of $O$, which is contained in $A$. Namely, let
$\widetilde{O}$ be the $N_1\times N_2$ upper left corner of $O$, with $N_1\le M_1$ and
$N_2\le M_2$. The average value of any function of matrix elements of $\widetilde{O}$ can
be computed either by integrating over $O$ or over $A$. In particular, the quantity
\be\label{I2U} \Wg^{O}_{N+1}(\beta)=\int_{\mathcal{O}(N+1)} dO \prod_{k=1}^n
\widetilde{O}_{k,j_{k}}\widetilde{O}_{k,j_{\widehat{k}}},\ee where $k\le N_1,N_2$ and the $j$'s only satisfy some matching of cosettype $\beta$, can also be written as
\be\label{I2}\Wg^{O}_{N+1}(\beta)=\frac{1}{\Y_2}\int_{AA^T< 1_{M_1}}
dA\det(1-AA^T)^{N_0/2} \prod_{k=1}^n A_{k,j_{k}}A_{k,j_{\widehat{k}}}.\ee

Notice that the right-hand-side of equation (\ref{I2}) is actually independent of $M_1$
and $M_2$.

\subsection{Sum over maps}

Analogously to the unitary case, we have \be \Wg^{O}_{N+1}(\beta)=\frac{1}{\Y_2}\int
d_G(A)e^{-\frac{N_0}{2}\sum_{q=2}^\infty\frac{1}{q}{\rm Tr}(AA^T)^q} \prod_{k=1}^{2n}
A_{k,j_{k}}A_{k,j_{\widehat{k}}},\ee where now $d_G(A)=e^{-\frac{N_0}{2}{\rm Tr}(AA^T)}$. The
diagrammatical considerations proceed as previously, except that we use the Wick's rule
of the real case and the resulting maps need not be orientable. Also, a map now
contributes $(-N_0/2)$ for each internal vertex and $1/N_0$ for each edge. This gives a
total contribution which is proportional to \be
N_0^{v-E}\left(-\frac{1}{2}\right)^v=\frac{N_0^\chi}{N^{F+\ell(\beta)}}\left(-\frac{1}{2}\right)^V(-2)^{\ell(\beta)},\ee
where $v$ is the number of internal vertices, $V=v+\ell$ is the total number of vertices,
$E$ is the number of edges and $\chi=F-E+V$ is the Euler characteristic, where $F$ is the
number of faces. When we take $M_2\to 0$, and then $M_1\to 0$, we arrive at maps with no closed boundary walks that avoid the marked vertices, having $2n$ faces, $n$ of each color. We thus arrive at the maps in the set $\mathcal{NB}(\beta,\chi)$.

The Gaussian normalization constant is
\begin{align} \Z_R=\int d_G(A)&=\mathcal{V}_2\int_0^\infty \prod_{i=1}^{M_1}
dt_ie^{-\frac{N_0}{2}t_i}t_i^{(M_2-M_1-1)/2}\prod_{j=i+1}^{M_1}|t_j-t_i|
\\&=\mathcal{V}_2\left(\frac{2}{N_0}\right)^{\frac{M_1M_2}{2}}\prod_{j=1}^{M_1}
\frac{\Gamma(1+j/2)\Gamma((M_2+1-j)/2)}{\Gamma(3/2)},\end{align} and we have \be
\lim_{M_2\to 0} \frac{\Z_R}{\Y_2}= \lim_{M_2\to
0}\left(\frac{2}{N_0}\right)^{\frac{M_1M_2}{2}}\prod_{j=1}^{M_1}\frac{\Gamma((N+2-j)/2)}{\Gamma((N-M_2+2-j)/2)}=1.\ee
Once again, the limit $M_1\to 0$ is trivial. This reduces $N_0$ to $N$. Taking into account the contribution of the maps, already discussed, we arrive at our Theorem 3.

\subsection{Sum over factorizations}

We now label the maps in $\mathcal{NB}(\beta,\chi)$ in order to relate them to
permutations. We only need to change slightly the labelling procedure we used for the
maps in $\mathcal{B}(\alpha,\chi)$ in Section 3.3. First, we replace the labels of the corners in dashed line by `hatted' versions. Second, instead of labelling corners, we now label half-edges, by rotating the previous labels counterclockwise. This is shown
in Figure 7 (where the hatted labels are enclosed by boxes, while the normal ones are
enclosed by circles).

The unhatted labels, read in anti-clockwise order around vertices, produce a permutation
$\Pi$ which is standard. This can be written as $\Pi=\pi\rho$, where $\pi\in S_n$ has
cycletype $\beta$ and the complement $\rho$ acts on the set $\{n+1,...,E\}$ where $E=n+m$ is the number of edges.
As before, $\rho$ has no fixed points, since all internal vertices of maps in $\mathcal{NB}(\beta,\chi)$ have even valence larger than 2.

A fixed-point free involution $\theta$ can be constructed from the
labels that appear at the ends of each edge. Namely, in the examples shown in Figure 7
it is given by $\theta=(1\widehat3)(\widehat13)(2\widehat4)(\widehat24)$ for a),
$\theta=(1\widehat3)(\widehat13)(25)(\widehat24)(\widehat46)(\widehat5\widehat6)$ for b) and
$\theta=(1\widehat3)(\widehat13)(25)(\widehat24)(\widehat4\widehat5)$ for c).

We also define the hatted version of any permutation $\pi$ by the equation
$\widehat\pi(a)=\pi^{-1}(\widehat a)$, assuming $\widehat{\widehat a}=a$. This is clearly an
involution. Permutations that are invariant under this operation are called
`palindromic', such as $(12\widehat 2 \widehat 1)$ or $(12)(\widehat 2 \widehat 1)$. Any permutation that
can be written as $\pi\widehat{\pi}$ where $\pi$ is another permutation is automatically
palindromic.

Define two special fixed-point free involutions, \be p_1=(1\,\widehat{1})(2\,\widehat{2})\cdots,\ee and
\be
p_2=(\widehat1\,2)(\widehat2\,3)\cdots(\widehat{\beta_1}\,1)(\widehat{\beta_1+1}\,\beta_1+2)\cdots(\widehat{\beta_1+\beta_2}\,\beta_1+1)\cdots.\ee
Notice that the cycles of $p_1$ contain labels which delimit corners of dashed line, while that the cycles of $p_2$ contain labels which delimit corners of solid line. Notice also that they factor the palindromic version of the vertex permutation, $p_2p_1=\Pi\,\widehat{\Pi}$.

\begin{figure}[tb]\label{figXO}
\includegraphics[scale=0.8,clip]{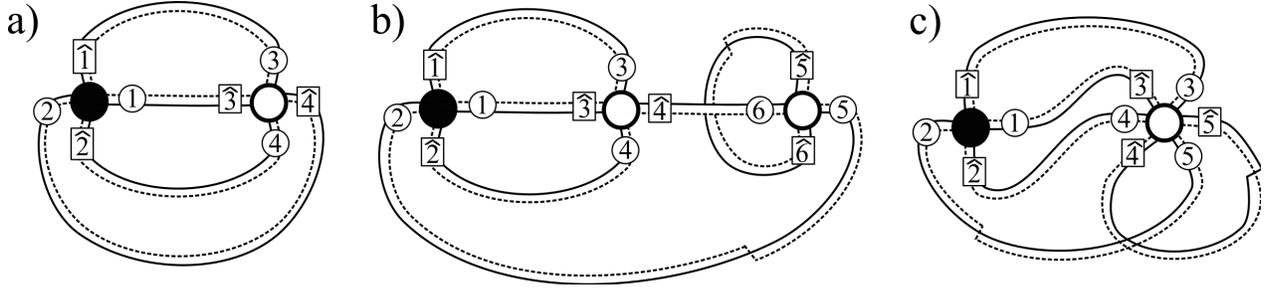}
\caption{Labelling some of the maps from Figures 1 and 2, in the way appropriate for the orthogonal
case.}
\end{figure}

By construction, the permutation $f_1=\theta p_1$ contains every other label encountered
along boundary walks around the faces delimited by boundaries in dashed line. For
example, in Figure 7 it would be $f_1=(13)(24)(\widehat4\widehat2)(\widehat3\widehat1)$ for a),
$f_1=(13)(246\widehat5)(5\widehat6\widehat4\widehat2)(31)$ for b) and
$f_1=(13)(24\widehat5)(5\widehat4\widehat2)(31)$ for c). In particular, $f_1$ is always
palindromic.

Conversely, permutation $f_2=p_2\theta$ contains every other label encountered along
boundary walks around the faces delimited by boundaries in solid line. In Figure 7 it
would be $f_2=(14)(23)(\widehat3\widehat2)(\widehat4\widehat1)$ for a), $f_2=(14)(2\widehat663)(\widehat3\widehat2)(\widehat4\widehat55\widehat1)$ for b) and $f_2=(14)(2\widehat43)(\widehat3\widehat2)(\widehat1\widehat55)$ for c).
Permutation $f_2$ needs not be palindromic. Notice that $f_1$ and $f_2$ are factors for
the palindromic version of the vertex permutation, $\Pi\,\widehat{\Pi}=f_2f_1$.

For the maps in $\mathcal{NB}(\beta,\chi)$ all boundary walks visit the marked vertices,
which means that all cycles of $f_1$ and of $f_2$ must have exactly one element in the
set $\{1,...,n,\widehat1,...,\widehat n\}$. Therefore, the permutations satisfy the conditions we
listed in Section 1.3.2. When we label the vertices of the map to produce a factorization, the same ambiguities arise as for the unitary group, which are accounted for by division by the factor $z_\rho$. We have therefore arrived at factorizations in $\mathcal{NF}(\beta,\chi)$ and proved our Theorem 4.

\section*{Appendix - Other factorizations}

\subsection{Unitary case}

A \emph{monotone} factorization of $\pi$ is a sequence  $(\tau_1,...,\tau_k)$ of transpositions $\tau_i=(s_i\,t_i)$, $t_i>s_i$ and $t_i\ge t_{i-1}$, such that $\pi=\tau_1\cdots\tau_k$. The number of transpositions, $k$, is the length of the factorization. Let $M_\alpha^k$ be the number of length $k$ monotone factorizations of $\pi$, with cycletype $\alpha$. Using the theory of Jucys-Murphy elements, Matsumoto and Novak showed that 
\be \Wg_N^U(\alpha)=(-1)^{n+\ell(\alpha)}\sum_{k=0}^\infty M_\alpha^k N^{-n-k}.\ee

A \emph{proper} factorization of $\pi$ is a sequence of permutations $(\tau_1,...,\tau_k)$, in which no one is the identity, such that $\pi=\tau_1\cdots\tau_k$. The \emph{depth} of a proper factorization is te sum of the ranks of the factors, $\sum_{j=1}^k r(\tau_j)$.

Let $P_\alpha^{k,d}$ be the number of proper factorizations of $\pi$, with cycletype $\alpha$, having length $k$ and depth $d$. It is known that \cite{stanleybook}
\be\label{frob} P_\alpha^{k,d}=\frac{1}{n!}\sum_{\lambda\vdash n}\chi_\lambda(1^n)\chi_\lambda(\alpha)\sum_{\mu_1\cdots\mu_k}\left\{\prod_{j=1}^k|C_{\mu_j}|\frac{\chi_\lambda(\mu_j)}{\chi_\lambda(1^n)}\right\}\delta_{\sum_jr(\mu_j),d}, \ee where all partitions $\mu$ are different from $1^n$. 
Starting from the character expansion of the Weingarten function,
\be \Wg_N^U(\alpha)=\frac{1}{n!^2}\sum_{\lambda\vdash n}\frac{\chi_\lambda(1^n)^2}{s_\lambda(1^N)}\chi_\lambda(\alpha),\ee Collins used \cite{collins} the Schur function expansion \be s_\lambda(1^N)=\frac{1}{n!}\sum_{\mu\vdash n}|C_\mu|\chi_\lambda(\mu)N^{\ell(\mu)}=\frac{\chi_\lambda(1^n)N^n}{n!}
\left(1+\sum_{\mu\vdash n,\mu\neq 1^n}|C_\mu|\frac{\chi_\lambda(\mu)}{\chi_\lambda(1^n)}N^{-r(\mu)}\right),\ee to arrive at the expression 
\be \Wg_N^U(\alpha)=\frac{1}{n!N^n}\sum_{\lambda\vdash n}\chi_\lambda(1^n)\chi_\lambda(\alpha)\sum_{k=1}^\infty(-1)^k\sum_{\mu_1\cdots\mu_k}
\prod_{j=1}^k|C_{\mu_j}|\frac{\chi_\lambda(\mu_j)}{\chi_\lambda(1^n)}N^{-r(\mu_j)}.\ee Comparing with (\ref{frob}) ones concludes that 
\be \Wg_N^U(\alpha)=\sum_{d=0}^\infty \left(\sum_{k=1}^d (-1)^kP_\alpha^{k,d}\right) N^{-n-d}.\ee

A \emph{cycle} factorization $\pi$ is a sequence of permutations $(\tau_1,...,\tau_k)$, in which all factors have only one cycle besides singletons, i.e. their cycletypes are hook partitions. \emph{Inequivalent cycle} factorizations are equivalence classes of cycle factorizations, two factorizations being equivalent if they differ by the swapping of adjacent commuting factors. Berkolaiko and Irving show \cite{BerkoIrv} that the number of such factorizations of $\pi$, with cycletype $\alpha$, having length $k$ and depth $d$, denoted by us $I_\alpha^{k,d}$, satisfy
\be \sum_{k}(-1)^kI_\alpha^{k,d}=\sum_{k}(-1)^kP_\alpha^{k,d}=(-1)^{n+\ell(\alpha)}M_\alpha^d.\ee These results are indexed by depth, but one can use Euler characteristic instead, by resorting to the equality $\chi=n+\ell(\alpha)-d$.

\subsection{Orthogonal case}

Consider again permutations acting on the set $[n]\cup\widehat{[n]}$. Let $h$ be the operation of `forgetting' the hat, i.e. $h(a)=h(\widehat a)=a$ for all $a\in[n]$.

Matsumoto defined the following analogue of monotone factorizations \cite{matsujack}. Let $\mathfrak{m}$ be a matching and let $(\tau_1,...,\tau_k)$ be a sequence of transpositions $\tau_i=(s_i\,t_i)$, in which all $t_i\in[n]$ with $t_i\ge t_{i-1}$ and $t_i>h(s_i)$, such that $\mathfrak{m}=\tau_1\cdots\tau_k (\mathfrak{t})$, where $\mathfrak{t}$ is the trivial matching. Let $\widetilde{M}_\alpha^k$ be the number of length $k$ such factorizations of some $\mathfrak{m}$ with cycletype $\beta$. Then,
\be \Wg_N^O(\beta)=\sum_{k=0}^\infty (-1)^k \widetilde{M}_\beta^k N^{-n-k}.\ee

The analogue of proper factorizations in this context is, for a permutation $\sigma$ with cosettype $\beta$, a sequence of permutations $(\tau_1,...,\tau_k)$, no one having the same cosettype as the identity, such that $\sigma=\tau_1\cdots\tau_k$. Let $\widetilde{P}_\beta^{k,d}$ be the number of such factorizations having length $k$ and depth $d$. We know from \cite{vassili,hanlon,jackson} that (actually these works only consider $k=2$, but the extension to higher values of $k$ is straightforward) \be\label{properO} \widetilde{P}_\beta^{k,d}=\frac{1}{(2n)!}\sum_{\lambda\vdash n}\chi_{2\lambda}(1^{2n})\omega_\lambda(\beta)\sum_{\mu_1\cdots\mu_k}\left\{\prod_{j=1}^k|K_{\mu_j}|\omega_\lambda(\mu_j)\right\}\delta_{\sum_jr(\mu_j),d},\ee where \be \omega_\lambda(\tau)=\frac{1}{2^nn!}\sum_{\xi\in H_n}\chi_{2\lambda}(\tau\xi)\ee are the zonal spherical function of the Gelfand pair ($S_{2n},H_n$) (they depend only on the cosettype of $\tau$; see \cite{donald}).

The relation between the above factorizations and orthogonal Weingarten functions comes as follows. The character-theoretic expression for the orthogonal Weingarten function is \cite{ColMat}
\be \Wg_N^O(\tau)=\frac{2^nn!}{(2n)!}\sum_{\lambda\vdash n}\frac{\chi_{2\lambda}(1^n)}{Z_\lambda(1^N)}\omega_\lambda(\tau),\ee where $Z_\lambda$ are zonal polynomials. Following the same procedure used for the unitary group, we expand $Z_\lambda(1^N)=\frac{1}{2^nn!}\sum_{\mu\vdash n}|K_\mu|\omega_\lambda(\mu)N^{\ell(\mu)}$ to arrive at \be \Wg_N^O(\tau)=\frac{2^nn!}{(2n)!}\sum_{\lambda\vdash n}\frac{\chi_{2\lambda}(1^n)\omega_\lambda(\tau)}{N^n}\sum_{k=1}^\infty (-1)^k\sum_{\mu_1\cdots\mu_k}
\prod_{j=1}^k\frac{|K_{\mu_j}|}{2^nn!}\omega_\lambda(\mu_j)N^{-r(\mu_j)}.\ee Comparing with (\ref{properO}), we see that 
\be \Wg_N^O(\tau)=\sum_{d=0}^\infty \left(\sum_{k=1}^d \frac{(-1)^k}{(2^nn!)^{k-1}}\widetilde{P}_\beta^{k,d}\right) N^{-n-d}.\ee 

Berkolaiko and Kuipers have provided a combinatorial description of the coefficients in the $1/N$ expansion of the function $\Wg_{N+1}^O$ \cite{GregJack1} (they actually worked with the so-called Circular Orthogonal Ensemble of unitary symmetric matrices, but the Weingarten function of that ensemble coincides \cite{matsucompact} with $\Wg_{N+1}^O$). A \emph{palindromic monotone factorization} is a sequence  $(\tau_1,...,\tau_k)$ of transpositions $\tau_i=(s_i\,t_i)$, with $t_i>s_i$ and $t_i\ge t_{i-1}$, such that $\pi\widehat{\pi}=\tau_1\cdots\tau_k\widehat{\tau_k}\cdots\widehat{\tau_1}$. Let $\widehat{M}_\beta^k$ be the number of length $k$ palindromic monotone factorizations of $\pi\widehat{\pi}$, with $\pi$ a permutation of cycletype $\beta$. Then,
\be \Wg_{N+1}^O(\beta)=\sum_{k=0}^\infty (-1)^k \widehat{M}_\beta^k N^{-n-k}.\ee  

An appropriate analogue of inequivalent cycle factorizations is currently missing, but we conjecture that, whatever they are, their counting function will be related to coefficients in $1/N$ expansions of orthogonal Weingarten functions.

\end{document}